\documentclass[pra,aps,showpacs,twocolumn,floatfix]{revtex4}
\usepackage{epsfig} \usepackage{graphics} \usepackage{bm}
\usepackage{amssymb}\usepackage{color}
\begin{document}
\title{Unconventional superfluidity in Bose-Fermi Mixtures}
\author{O. Dutta$^1$, M. Lewenstein$^{1,2}$}
\affiliation{${}^1$ ICFO-Institut de Ci{\`e}ncies Fot{\`o}niques, Mediterranean Technology Park, 08860 Castelldefels (Barcelona), Spain,}
\affiliation{${}^2$ ICREA-Instituci{\`o} Catalana de Recerca i Estudis Avan\c{c}ats, Lluis Companys 23, 08010 Barcelona, Spain.}

\date{\today}

\begin{abstract}

\end{abstract}

\maketitle

\textbf{Pairing between fermions that attract each other, reveal itself to the macroscopic world
in the form of superfluidity. Since the discovery of fermionic superfluidity \cite{sup},
intense search has been going on to find various unconventional forms of fermion pairing
\cite{am} as well as to increase the transition temperature \cite{htc}.
Here, we show that a two dimensional mixture of single-component fermions and dipolar
bosons allows to reach experimentally feasible superfluid
transition temperatures for non-standard pairing symmetries.
Excitations in these superfluids are anyonic and their statistics depends on the order of their
permutations, i.e is non-Abelian. Our results provide for the first time an example of a highly tunable system
 which exhibits various kind of pairing symmetry and
high transition temperature. Additionally, they provide a playground to
observe anyonic excitations and their braiding properties.}

Attraction between fermionic particles favours pairing of fermions
resulting in superfluidity of the system. The paired fermions, known
as Cooper pairs, can have different kind of internal symmetries. The
common ones found in nature have $s$-wave and $d$-wave like
internal structure and conserves parity and time reversal symmetry.
Also, Cooper pairs with chiral $p_x+i p_y$-wave internal structure have
been proposed for the observed superfluidity of electrons in
Strontium Ruthnate \cite{str}. This kind of pairing breaks the time-reversal
symmetry. Spin-less chiral $p$-wave superfluid state has formal
resemblance with the ``Pfaffian" state proposed in relation to the fractional
quantum Hall state with filling factor $5/2$ \cite{QH1, CN1, read1}. When
confined in a two dimensional geometry, excitations in the chiral $p$-wave superfluid become
non-abelian anyons. Anyons are particles living in a two dimensional plane that under exchange
 do not behave as bosons or fermions. In some cases, exchange of two such
particles depends on the order of the exchange \cite{W1, W2, W3}.
Particles obeying such laws are called non-Abelian anyons. Apart
from fundamental interest on the occurrence of such particles, non-Abelian anyons find remarkable
applications in the field of quantum information for quantum memories and fault-tolerant quantum
computation \cite{K1}. Recently it has been shown that quasi-particles living in stable
vortex excitations of chiral two-dimensional p-wave spinless superfluids obey non-Abelian statistics
\cite{iva, CN2}. Using $p$-wave Feshbach resonances in fermionic
ultracold atoms, such superfluids can be realized in principle, but this procedure is very difficult because
of non-elastic loss processes \cite{gur1}.

        Bose-Fermi mixtures are another candidate for creating superfluidity
in fermions via boson mediated interactions and have
formal resemblance with phonon mediated superconductivity in metals \cite{sup}.
It was found, however, when the bosons and single-component fermions are completely
mixed, the maximum possible transition temperature is of the order
of $10^{-5}$T$_F$ for $p$-wave pairing, where T$_F$ is the Fermi
Temperature \cite{viv1}. Any attempt to increase the transition
temperature by increasing the boson-fermion interaction strength or
fermionic density results in phase separation between the mixture.
Hence it is experimentally hardly possible to study this
phenomenon for unconventional superfluidity in ultracold atomic systems.

             Here, we show a possible way of overcoming these difficulties.
We study the property of superfluidity in Bose-Fermi mixtures,
where bosons are interacting via long-ranged dipolar interactions.
We show that the transition temperature for $p$-wave superfluidity can
become comparable to the Fermi energy. We find that other more exotic Cooper pairs
with $f$- and $h$-wave internal symmetries are possible in certain range of Fermi energies without bosons
and fermions separating. To the best of our knowledge, for the first time a system is proposed
where conventional pairing mechanism gives rise to different exotic internal structures of the Cooper pairs
with strong interaction in respective angular momentum channels. In addition, we study the excitations in chiral states
of the odd-wave superfluids and point out their non-Abelian anyonic nature.

             Experimentally, an available bosonic species,
where prominent dipolar interaction can be achieved using
Feshbach resonance is Cr$^{52}$ \cite{Pf1, Pf2}. Another route towards
achieving dipolar condensate is to experimentally realize quantum
degenerate heteronuclear molecules \cite{Jin1} which have permanent
electric moment. Thus in the near future a quantum
degenerate mixture of dipolar bosons and fermions will be achievable
experimentally.

\section{SYSTEM}
\subsection{Dipolar Bose Condensate}

\begin{figure*}[ht]
\begin{center}
$\begin{array}{c@{\hspace{1in}}c} \multicolumn{1}{l}{\mbox{\bf (a)}}
&
    \multicolumn{1}{l}{\mbox{\bf (b)}} \\ [-0.53cm]
\epsfig{file=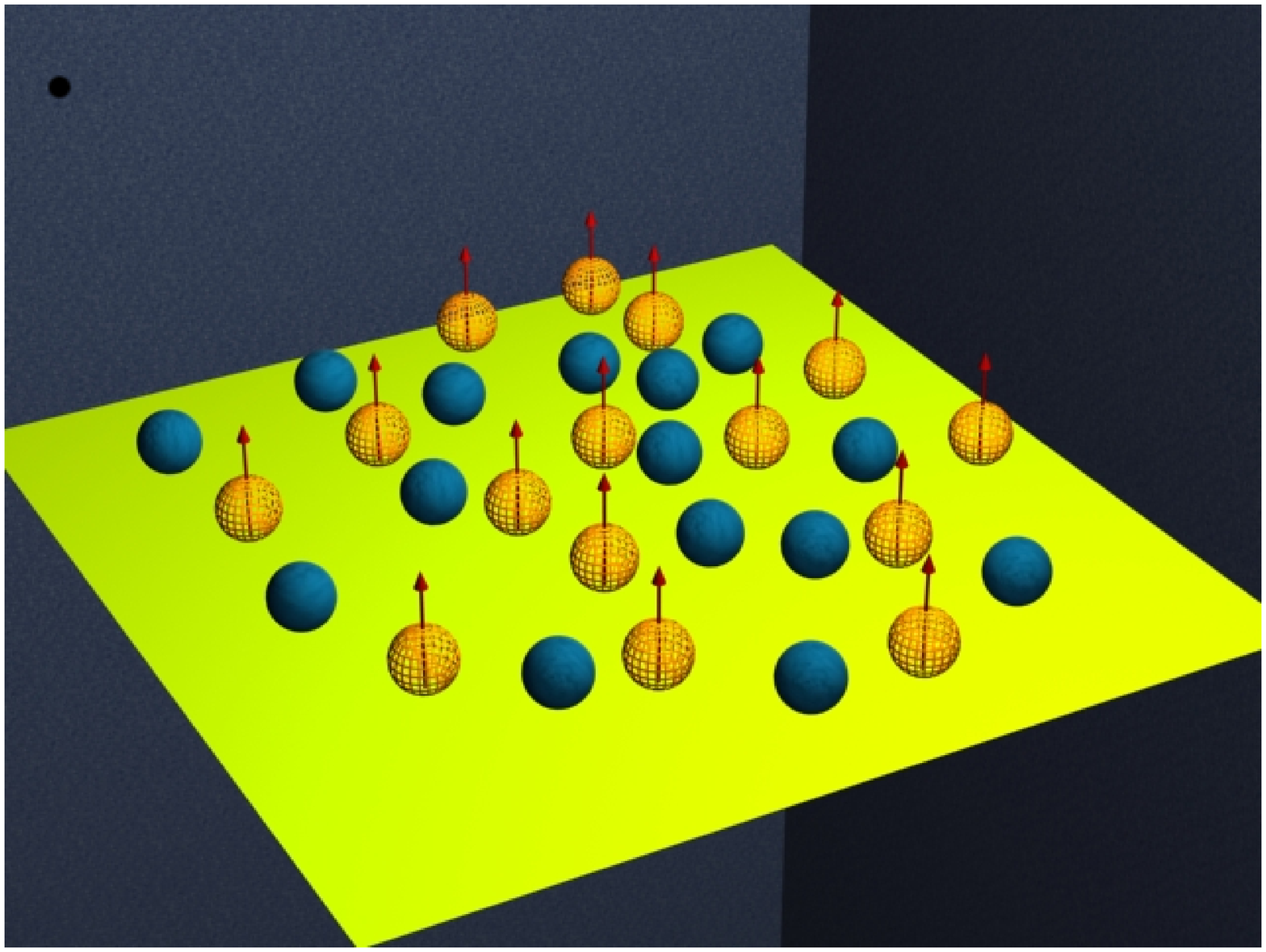, width=5.5cm, height=4.5cm} &
    \epsfig{file=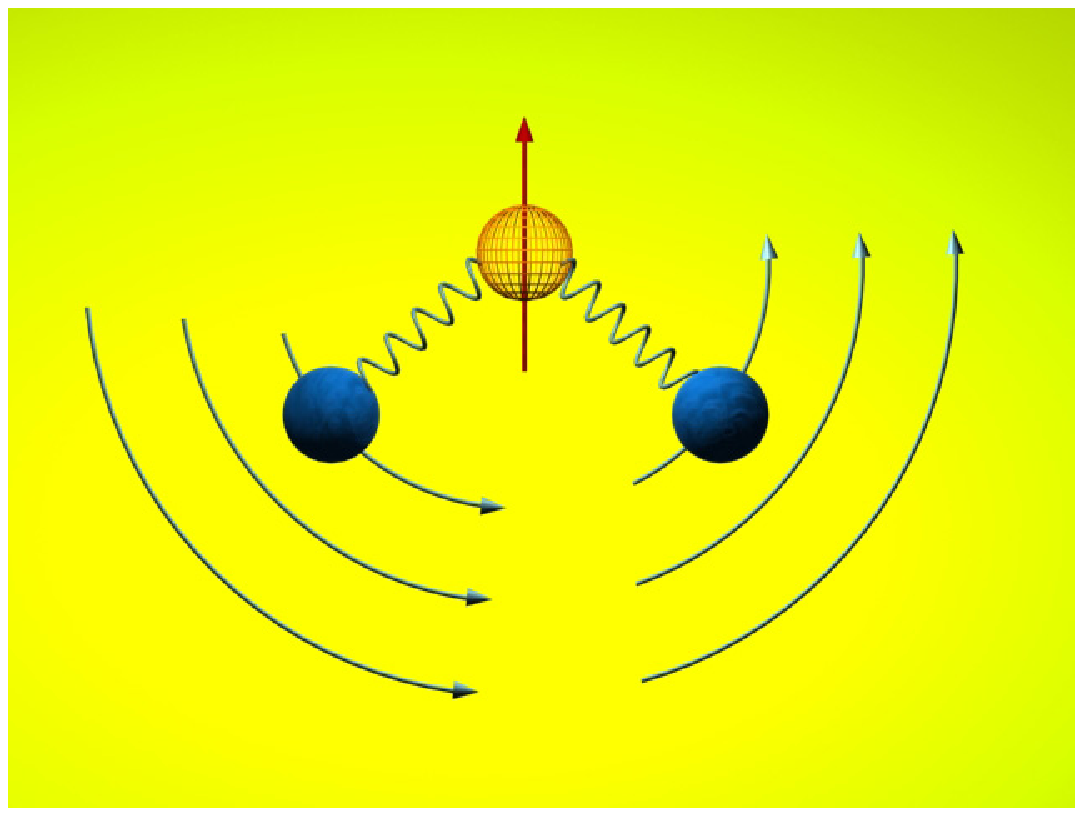, width=5.5cm, height=4.5cm} \\ [0.4cm]
\mbox{\bf (a)} & \mbox{\bf (b)}
\end{array}$
\caption{\label{fig1}a) Schematic image of a two-dimensional mixture of dipolar bosons (semi-transparent spheres with arrows)
and fermions (blue spheres). b) The interaction between the bosons and fermions induces interaction between the fermions. This
results in superfluidity of the fermions with various angular momenta as inducated by the arrows around the fermions.}
\end{center}
\end{figure*}

Our system, as sketched in Fig.\ref{fig1}a, consists of dipolar bosons mixed with single component
fermions confined in a quasi-two dimensional geometry by a harmonic
potential with frequency $\omega_z$ and oscillator length $\ell_0$. Here we present a brief
overview of dipolar condensates with a focus on our present
problem (for details see methods). First, we assume that the bosons are polarized along the $z$
direction. The dipolar interaction reads $
V_{\rm dd}=\frac{4\pi g_{\rm dd}}{3}( 3k^2_z/k^2-1 )$ in momentum space, where $g_{\rm
dd}$ is the dipole-dipole interaction strength. For atoms $g_{\rm
dd}= \mu_0 \mu^2_m/4\pi $, and for dipolar molecules $g_{\rm
dd}=\mu^2_e/4\pi\epsilon_0 $ where $\mu_m$ and $\mu_e$ are the
magnetic moment of the atoms and the electric dipole moment of the
molecules, respectively. We
assume that the $z$ dependance of bosonic wavefunction is given by a
Thomas-Fermi profile with radius $R_z$. By integrating out the $z$ direction,
the effective dipolar interaction takes the form $V_{\rm
eff}=\frac{g_{\rm dd}}{R_z} \mathcal{V}(k_{\bot})$, where
$k^2_{\bot}=k^2_x+k^2_y$ and $\mathcal{V}(k_{\bot})$ denotes the
shape of the interaction. Next, we define a dimensionless dipolar
interaction strength $g_{\rm 3d}=8\pi m_b g_{\rm dd} n_b
\ell_0/5\hbar^2$ which will be used later, where $\ell_0$ is the ground state oscillator length.
$\mathcal{V}(k_{\bot})$ is repulsive for small
momentum and attractive in the high momentum limit.
 At low temperature, the ground state of the bosons is a condensate with
fluctuations around this state. Generally, the spectrum of these
excitations, denoted by $\Omega(\vec{k_{\bot}})$, can be divided
into two parts: i) $\sim k_{\bot}$, phonon spectrum for small
momenta and ii) $\sim k^2_{\bot}$, free-particle like spectrum for
higher momenta \cite{pit}. For the dipolar condensates and $g_{\rm
3d}$ greater than a critical value, $\Omega(\vec{k_{\bot}})$ has a
minimum at momentum $\tilde{k}_0$ \cite{Lew1}, where $k_0$ is in
intermediate momentum regime as shown in Fig.~2b. Following Landau,
the excitations around the minimum are called ``rotons''. With
increasing $g_{\rm 3d}$ the excitation energy at $k_0$ decreases and
eventually vanishes for a critical particle density as shown in
Fig.~2b. When the particle density exceeds that critical value, the
excitation energy becomes imaginary at finite momentum and the
condensate becomes unstable. The ``roton" part of the spectrum is
absent in condensates with contact interaction.

\subsection{ Effective interaction}
In this section we discuss the effect of bosons on fermions. The
fermions are interacting with the bosons via short ranged contact
interaction of strength $g_{\rm bf}$. The fermionic density dependance
on $z$ is given by a normalized Gaussian with width
$\ell_f$. We also define the $dimensionality$ parameter for the fermions as
$\eta=\mu/\hbar\omega_z$, where $\mu$ is the chemical potential for
the fermions. For $\eta<1$ the fermionic system is quasi-two
dimensional, and we can assume $\ell_f \sim \sqrt{\hbar/m_f
\omega_z}$, i.e the width is given by the oscillator length of the
trapping potential. $\eta>1$ corresponds to three dimensional
fermionic system and $\ell_f$ is given by minimizing the
Thomas-Fermi energy functional, $\ell_f \sim \eta^{1/4} (\hbar/m_f
\omega_z)^{1/2}$. The
fluctuations in fermionic density couple to the density
fluctuations present in the Bose condensate. Due to the momentum dependance of
the bosonic excitations, the condensate-fermion interaction becomes function of momentum. Integrating out the
bosonic degree of freedom results in effective interaction between
the fermions \cite{wang1},
\begin{equation}\label{vph}
V_{\rm ph}(\vec{q_{\bot}}, \omega)= \frac{9g^2_{\rm bf}\alpha^2}{16\pi R^2_z}
 \frac{n_b q^2_{\bot}/m_b}{\omega^2-\Omega^2(\vec{q_{\bot}})},
\end{equation}
where $\vec{q}^2_{\bot}=2k^2_f(1-\cos\phi)$, is the momentum exchange between the
interacting particles along the Fermi surface, and $\alpha$ is a number which depends on $R_z/\ell_f$ (see Methods).
This form of interaction has formal
resemblance to superconductivity in metals. Due to the momentum
dependance, if attractive, $V_{\rm ph}(\vec{q_{\bot}}, \omega)$
makes the Fermi surface unstable against formation of Cooper pairs
with higher internal symmetries. Assuming momentum transfer around
Fermi momentum $k_f$, in two-dimensions we can expand Eq.~(\ref{vph})
as
$$
V_{\rm ph}(\vec{k_{\bot}}, 0)= \frac{3g^2_{\rm bf}N_0}{8\pi g_{\rm
dd}\mathcal{V}(0)\ell_0} \sum_{m=...,-1,0,1,...}\lambda_m e^{im\phi},
$$
where the dimensionless effective interaction between the fermions
in angular momentum channel $m$ is given by
\begin{eqnarray}\label{lam1}
\lambda_m &=&\alpha^2\int^{2\pi}_0 \frac{\exp [ i m \phi ] d\phi/2\pi}
{\frac{\eta R^2_z}{g_{\rm 3d} \ell^2_f}(1-\cos \phi)+\frac{R^2_z}{\ell^2_0}\mathcal{V}(\sqrt{\frac{R_z}{{\ell_f}}}\eta(1-\cos\phi))}, \nonumber\\
\end{eqnarray}
and $N_0=m_f/\pi^2\hbar^2$ is the two-dimensional density of states
for the fermions. In the present paper, $\lambda_m>0$ denotes
attractive interaction. As we are considering single component
fermions, pairing will occur in the odd angular momentum channels.
The effective dipole-dipole interaction between the bosons is given
by the density dependent term $g_{\rm 3d}$. The expression for
effective interaction obviously makes sense as long as the
excitation frequencies of the bosons are real. To look into the
properties of effective interaction, we consider first the
interaction strength in $p$-wave channel $\lambda_1$.
\begin{figure*}[ht]
\begin{center}
$\begin{array}{c@{\hspace{1in}}c} \multicolumn{1}{l}{\mbox{\bf (a)}}
&
    \multicolumn{1}{l}{\mbox{\bf (b)}} \\ [-0.53cm]
\epsfig{file=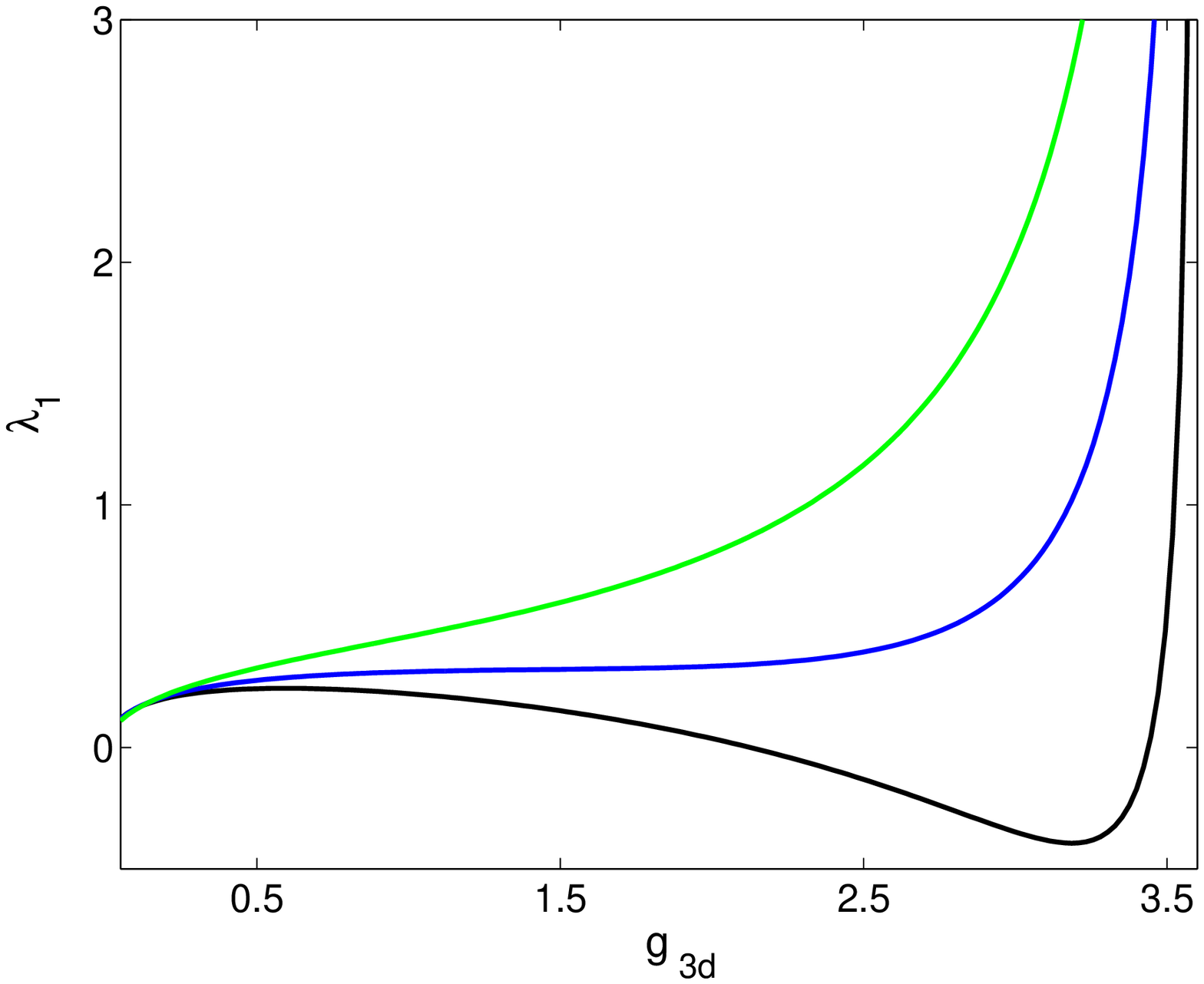, width=5.5cm, height=4.5cm} &
    \epsfig{file=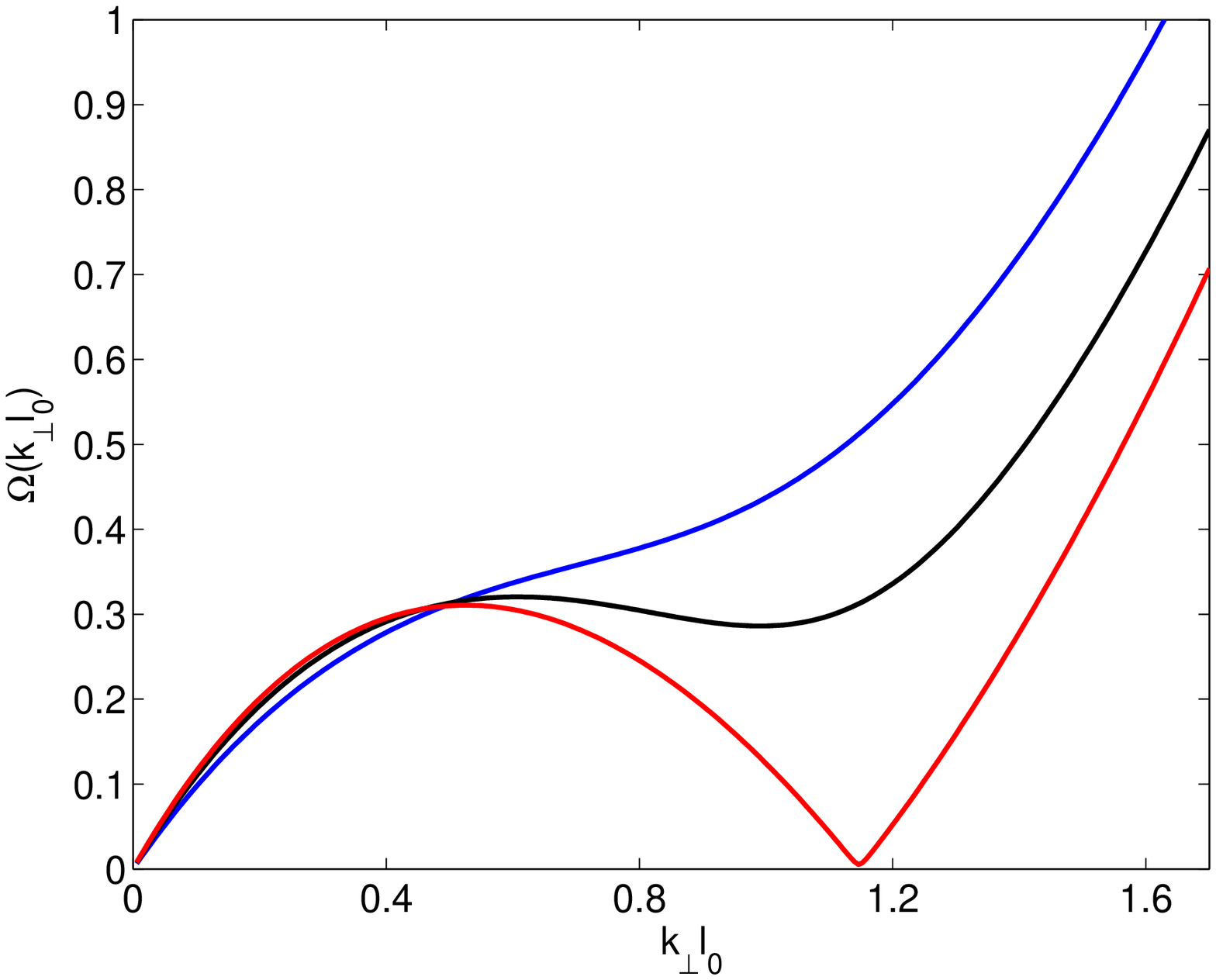, width=5.5cm, height=4.5cm} \\ [0.4cm]
\mbox{\bf (a)} & \mbox{\bf (b)}
\end{array}$
\caption{\label{fig2}a) Effective interaction strength in
the $p$-wave channel $\lambda_1$ as a function of boson-boson
effective interaction strength $g_{\rm 3d}$. The green, blue and black
lines correspond to $\eta=0.50, .60, .90$ respectively. b) The excitation spectrum
$\Omega(k_{\perp})$ of the two dimensional Bose gas plotted
as a function of $k_{\perp}$. The blue, black and red lines
correspond to $g_{\rm 3d}=2.0, 3.0, 3.61,$ respectively.}
\end{center}
\end{figure*}
As shown in Fig.~2b, increasing $g_{\rm 3d}$ results in smaller roton
gap, and when $g_{\rm 3d} \sim 3.61$, the roton minimum touches the
zero energy axis. Subsequently the effective fermionic interaction
$\lambda_1$ increases and diverges as $g_{\rm3d} \rightarrow 3.61$,
as shown in Fig.~2a. Depending on the dimensionality $\eta$, the
rate of divergence changes and the interaction can become repulsive. This Feshbach
resonances-like characteristics are the novel feature of this system.

\begin{figure}[ht]
\begin{center} \epsfig{file=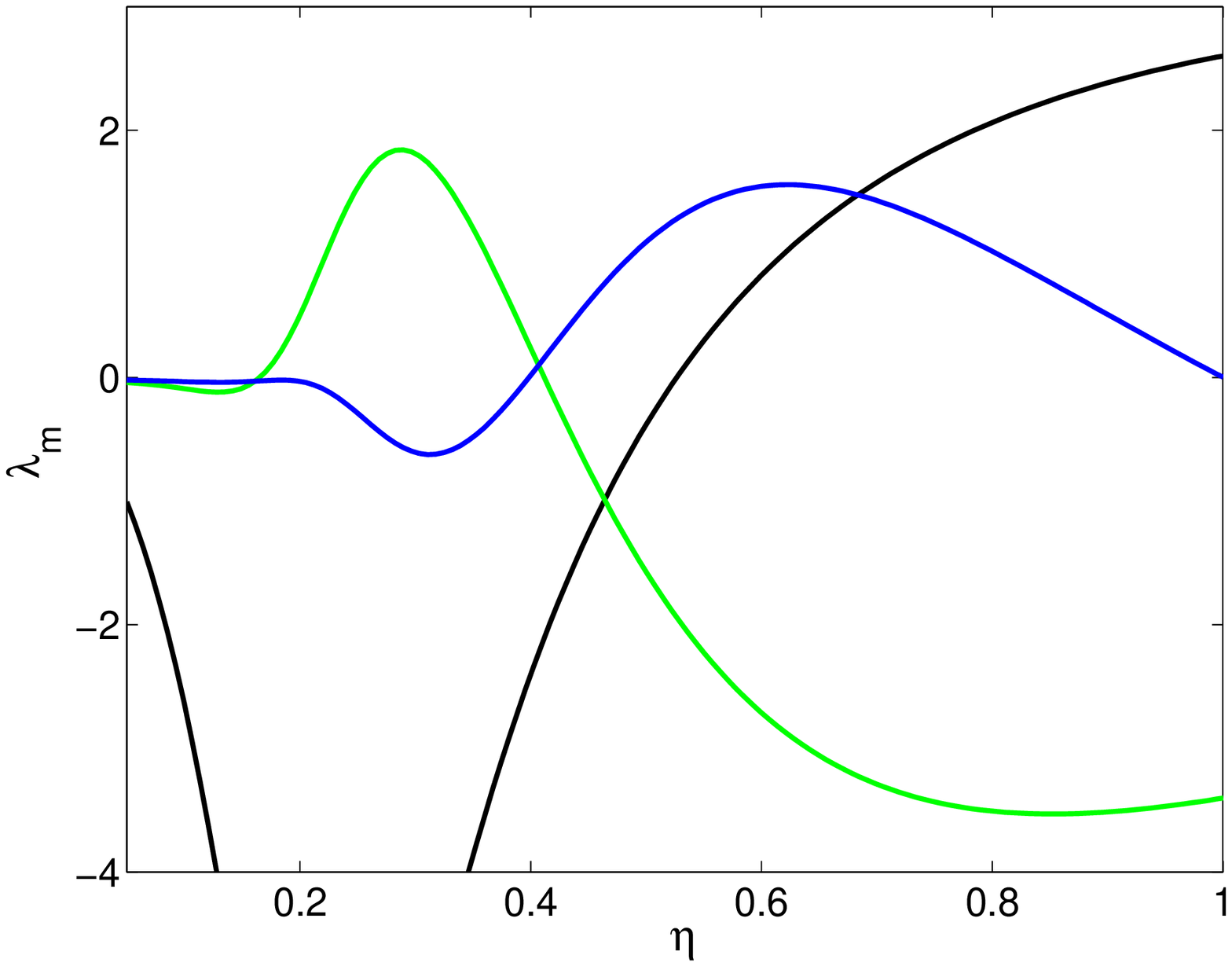,width=5.5cm, height=4.5cm} \caption{\label{fig3}
Figure of effective interaction strength in the channel $m=1$(black
line),$=3$(green line),$=5$(blue line) as a function of the fermion
dimensionality parameter $\eta$. We fixed the bosonic effective
interaction strength at $g_{\rm 3d}=3.1$. The negative value of the
interaction reflects it's repulsive nature.}
\end{center}
\end{figure}

Next we look into the variation of $\lambda_m$ as a function of
$\eta$ as depicted in Fig.~\ref{fig3}. Here for concreteness we
consider the particular case corresponding to a Chromium-Potassium
mixture with $m_b=52$a.m.u. and $m_f=40$ a.m.u.. Using these masses,
the interaction strengths in the channels
$m=1$(p-wave),$=3$(f-wave),$=5$(h-wave) have been plotted in Fig.~3.
As $\eta \rightarrow 1$, interaction in the $p$-wave channel becomes
predominant. This trend is also followed when we are inside the three dimensional
limit with $\eta > 1$. But, with decreasing dimensionality we find
that the predominant interaction channel changes surprisingly from
$m=1$ to $m=5$, the $h$-wave channel. Then $\lambda_5$ goes through a
maximum attraction around $\eta \sim .6$ and the interaction strength in the $p$-wave channel
becomes repulsive. In this region, the repulsive $p$-wave
interaction will renormalize the mass of the fermions and the
superfluid instability will be due to $m=5$ channel. By decreasing
$\eta$ further, both $\lambda_{1,5}$ become repulsive whereas the
$f$-wave channel with $m=3$ becomes attractive. Decreasing $\eta$
further results in a situation, where all the channel have negligible
interaction. The interaction strength at individual channel can be
increase further by increasing the bosonic density $n_b$ closer to
the critical value.

          Another obvious contribution to the strength of the interaction is the boson-fermion
contact interaction strength $g_{\rm bf}$. The condition for
mechanical stability of the Bose-Fermi mixtures reads \cite{viv1}
$$
\frac{3g^2_{\rm bf}N_0}{8\pi g_{\rm
dd}\mathcal{V}(0)\ell_0}\alpha^2 < 1,
$$
which constraints the magnitude of $g_{\rm bf}$. From Fig.~2a and
Fig.~3 it is clear that the maximum attractive interactions in each
channel can be $\lambda_{1},\lambda_{2}, \lambda_{3}> 1$ which are
within the strong-coupling regime. Consequently maximum transition
temperatures possible in all three channels are close to the fermi
temperature. At this point let us compare this to the case of usual
boson-fermion mixture, where highest interaction strength possible
is $\lambda_1 \sim .1$ \cite{viv1, wang1}. The maximum transition
temperature possible is for p-wave symmetry and $T_c$ is of the
order $\sim \exp(-1/.1)\mu \sim 10^{-5}\mu$. Note, then we have an
improvement in $T_c$ of the order of $10^5$.

\section{Chirality and non-Abelian Anyons} By considering the
superconducting gap equation at low temperature, the gap is
maximum when the order parameter breaks time reversal symmetry
\cite{am}. From now on we assume that the order parameters are
denoted by
$$
\Delta_m=\Delta_0(\vec{r}) \left [ \frac{k}{k_f} \right ]^m e^{i m
\theta}
$$
where $k_x=k\cos \theta$, $k_y=k\sin \theta$ and $\Delta_0(\vec{r})$
is the center of mass amplitude of the Cooper pairs with $\vec{r}$
being the center of mass coordinate of the pair. For vortex state
$\Delta_0(\vec{r})$ is defined as: i) $\Delta_0(\vec{r})=0, r<\xi$
and ii) $\Delta_0(\vec{r})= \Delta_0 \exp(i\phi), r\geq\xi$, where
$r=\sqrt{x^2+y^2}$, $\tan \phi=y/x$. $\xi$ is the size of the core of the vortex.
The vortex state of the $p$-wave
superfluids always has a zero-energy bound quasi-particle state
\cite{vol1, CN1, gur2}. Now we discuss the asymptotic solutions for the zero-energy
bound state for $f$- and $h$-wave order
parameters. The quasi-particle states in a single vortex can be
found in the limit of large distance from the vortex core by solving
the Bogoliubov-DeGennes equation,
\begin{widetext}
\begin{eqnarray}\label{bdg}
H_0 u_m+(-i)^m \frac{\Delta_0}{k^m_f} e^{i\phi/2}
\left [ e^{-i\phi} \left ( \frac{
\partial }{\partial r} - \frac{i}{r}\frac{ \partial
}{\partial \phi} \right ) \right ]^m e^{i\phi/2} v_m &=& E u_m \nonumber\\
- H_0 v_m  + (i)^m \frac{\Delta_0}{k^m_f} e^{-i\phi/2} \left [
e^{i\phi} \left ( \frac{
\partial }{\partial r} + \frac{i}{r}\frac{ \partial
}{\partial \phi} \right ) \right ]^m e^{-i\phi/2} u_m &=& E v_m,
 \nonumber\\
\end{eqnarray}
\end{widetext}
where $E$ is the energy of the quasi-particles denoted by $u_m,v_m$. We
particularly look for zero energy solutions with bounded $u_m,v_m$
with the property $u_m=v^{*}_m$ \cite{gur2}.
 The asymptotic solution of Eq. (\ref{bdg}) with different
orbital symmetries read,
\begin{equation} \label{angtop}
\bf \left[
\begin{array}{c}
u_1  \\
           \\
u_3  \\
                 \\
u_5               \\
\end{array}
\right] \sim \bf \left[
\begin{array}{c}
\exp \left ( -\frac{\Delta_0}{v_f} r \right ) \\
\exp \left ( -\frac{m^2_f v_f^3}{6 \Delta_0} r \right ) e^{2i\phi} \\
\exp \left ( - \left [ \frac{m^4_f v_f^5}{2\Delta_0} \right ]^{1/3}
r \right ) e^{4i\phi} \\
\end{array}
\right].
\end{equation}
The zero-energy solution for each odd-wave parameter corresponds to
different angular momentum channel of the quasi-particles inside a
vortex core. These results can also be carried out by applying the
``index theorem" \cite{tew}. For temperature smaller than the energy gap
$\Delta^2_0/\mu$, only the zero energy mode is occupied. The quasi-particle operator in that situation
is written as $\gamma_m=\int d^2r (u_m(r)c^{\dagger}(r) + v_m(r)c(r))$, which acts as
Majorana fermion \cite{iva, CN1, gur2}.  $\gamma_n$ obeys non-Abelian
statistics and can be used for quantum computing \cite{CN2}. In
order to perform quantum computational task, existence of several well separated vortices is necessary.
From Eq.~(\ref{angtop}) we find
that the condition for non-overlapped states can be achieved in
$m=3$ states with higher distance between the vortices than the case
for p-wave for similar values of gap and fermi energy.

\section{Acknowledgement}
This work is financially supported by the Spanish MEC QOIT (Consolider Ingenio 2010) projects,
TOQATA (FIS2008-00784), MEC/EST project FERMIX (FIS2007-29996-E), and ERC advanced grant QUAGATUA.

\section{Methods}

First we discuss the Hamiltonian describing the bosonic system which is homogeneous in the $x-y$ plane
and trapped in the $z$ direction by a harmonic potential with frequency $\omega_z$.
In the considered regime of parameters the bosonic density $n_b(x,y,z)$ is given by
$$
n_b(x,y,z)=\frac{3n_b}{4R_z} \left (1-\frac{z^2}{R^2_z} \right ),
$$
where the Thomas-Fermi radius $R_z$ is determined variationally. By minimizing the
mean field energy of the Bose condensate within Thomas-Fermi regime we find that
$R_z/\ell_0=(5g_{\rm 3d}/2)^{1/3}$. After integrating over $z$ dependance of the density profile of bosons,
the dipolar interaction takes the form $V_{\rm eff}=\frac{3 \pi g_{\rm dd}}{2R_z}
\mathcal{V}(\tilde{k}_{\bot})$ where
\begin{eqnarray}\label{veff}
\mathcal{V}(\tilde{k}_{\bot})&=&\frac{1}{\tilde{k}^5_{\bot}} \left [ 4\tilde{k}^3_{\bot} - 6\tilde{k}^2_{\bot}
- 6(1+\tilde{k}^2_{\bot})\exp(-2\tilde{k}_{\bot})+6 \right ]\nonumber\\
&-&\frac{8}{15}+\frac{2}{5\pi}\frac{g}{g_{\rm dd}},\nonumber
\end{eqnarray}
and $\tilde{k}_{\bot}=k_{\bot}R_z$ and $g$ is the contact interaction between the bosons which is assumed to be negligible
in our case.
Subsequently we write the Hamiltonian of the dipolar bosons in the
condensed phase, $H_b = \sum_{\vec{k_{\bot}}} \Omega(\vec{k_{\bot}})
b^{\dagger}_{\vec{k_{\bot}}} b_{\vec{k_{\bot}}}$, where
$b^{\dagger}_{\vec{k_{\bot}}}$ and $b_{\vec{k_{\bot}}}$ are
Bugoliubov operators. The excitation spectrum is given in the units
of trap frequency,
$$\label{exc}
\Omega^2({k}_{\bot}\ell_0)=\frac{[k_{\bot}\ell_0]^4}{4}+ g_{\rm 3d}\frac{\ell_0}{R_z}
\mathcal{V}\left ( {k}_{\bot}\ell_0 \frac{R_z}{\ell_0} \right ) [k_{\bot}\ell_0]^2.
$$

           Next, we consider the Hamiltonian describing the fermions and the boson-fermion interaction.
Kinetic energy for the single component non-interaction fermions, is
characterized by the Hamiltonian $H_f = \sum_{\vec{k_{\bot}}} \left
[ \epsilon_f(\vec{k_{\bot}}) - \mu \right ]
c^{\dagger}_{\vec{k_{\bot}}} c_{\vec{k_{\bot}}}$, where
$c^{\dagger}_{\vec{k_{\bot}}}$ and $c_{\vec{k_{\bot}}}$ are
fermionic creation and destruction operator.
$\epsilon_f(\vec{k_{\bot}}) = \vec{k_{\bot}}^2/2m_f $ is the
dispersion energy of the fermions with mass $m_f$. The density
profile of fermions along the $z$ direction is approximated by a
gaussian with width $\ell_f$.

Including the fluctuations in the Bose condensate in the $x-y$
plane, the condensate-fermion interaction Hamiltonian can be written
as
\begin{eqnarray}\label{hambf}
H_{\rm bf} &=& \frac{3g_{\rm bf}}{4\sqrt{\pi}
R_z}\alpha\sum_{\vec{k_{\bot}}, \vec{q_{\bot}}} \gamma(\vec{k_{\bot}})
c^{\dagger}_{\vec{k_{\bot}}} c_{\vec{q_{\bot}}-\vec{k_{\bot}}} \nonumber\\
&& \left [ b_{\vec{k_{\bot}}} + b^{\dagger}_{-\vec{k_{\bot}}} \right
] ,\nonumber
\end{eqnarray}
where the momentum
dependent coupling constant is given by
$\gamma(\vec{k_{\bot}})=\sqrt{2n_b\epsilon_b(\vec{k_{\bot}})/\Omega(\vec{k_{\bot}})}$ and
$$
\alpha=\frac{\ell_f}{R_z}\exp \left (-\frac{R^2_z}{\ell^2_f} \right )
- \frac{\sqrt{\pi}}{2} \left ( \frac{\ell^2_f}{R^2_z}-2 \right ) {\rm erf} \left ( \frac{R_z}{\ell_f} \right ),
$$
with ${\rm erf}(..)$ being the error function.

\end{document}